\documentclass[aps,pre,reprint,floatfix]{revtex4-2}
\usepackage[utf8]{inputenc}
\usepackage[T1]{fontenc}

\usepackage{amsmath}
\usepackage{amssymb}
\usepackage{amsfonts}
\usepackage{graphicx}
\usepackage{listings}
\usepackage{siunitx}
\usepackage{xcolor}
\usepackage{physics} % upright d with \dd

\usepackage[colorlinks=true,citecolor=blue,urlcolor=red]{hyperref}

\newcommand{\W}{\mathcal{W}}
\newcommand{\Q}{\mathcal{Q}}
\newcommand{\V}{V}
\newcommand{\K}{K}

\newcommand{\Teff}{T_{\textrm{eff}}}

\definecolor{codegreen}{rgb}{0,0.6,0}
\definecolor{codegray}{rgb}{0.5,0.5,0.5}
\definecolor{codepurple}{rgb}{0.58,0,0.82}
\definecolor{backcolour}{rgb}{0.95,0.95,0.92}

\lstdefinestyle{pythoncodestyle}{
    backgroundcolor=\color{backcolour},   
    commentstyle=\color{codegreen},
    keywordstyle=\color{magenta},
    numberstyle=\tiny\color{codegray},
    stringstyle=\color{codepurple},
    basicstyle=\ttfamily\footnotesize,
    breakatwhitespace=false,         
    breaklines=true,                 
    captionpos=b,                    
    keepspaces=true,                 
    numbers=left,                    
    numbersep=5pt,                  
    showspaces=false,                
    showstringspaces=false,
    showtabs=false,                  
    tabsize=2
}

\newcounter{AppendixEquation}
\makeatletter
\@addtoreset{equation}{AppendixEquation}
\makeatother

\newcounter{AppendixFigure}
\makeatletter
\@addtoreset{figure}{AppendixFigure}
\makeatother

\begin{document}

\title{Dynamics of information erasure and extension of Landauer's bound to fast processes}
\author{Salamb\^{o} Dago}
\author{Ludovic Bellon}
\email{ludovic.bellon@ens-lyon.fr}
\affiliation{Univ Lyon, ENS de Lyon, CNRS, Laboratoire de Physique, F-69342 Lyon, France}

\begin{abstract}
Using a double-well potential as a physical memory, we study with experiments and numerical simulations the energy exchanges during erasure processes, and model quantitatively the cost of fast operation. Within the stochastic thermodynamics framework we find the origins of the overhead to Landauer's Bound required for fast operations: in the overdamped regime this term mainly comes from the dissipation, while in the underdamped regime it stems from the heating of the memory. Indeed, the system is thermalized with its environment at all time during quasi-static protocols, but for fast ones, the inefficient heat transfer to the thermostat is delayed with respect to the work influx, resulting in a transient temperature rise. The warming, quantitatively described by a comprehensive statistical physics description of the erasure process, is noticeable on both the kinetic and potential energy: they no longer comply with equipartition. The mean work and heat to erase the information therefore increase accordingly. They are both bounded by an effective Landauer's limit $k_B \Teff \ln{2}$, where $\Teff$ is a weighted average of the actual temperature of the memory during the process.
\end{abstract}

\maketitle

Information processing in the physical world comes with an energetic cost: erasing a 1-bit memory at temperature $T_0$ requires at least $k_BT_0 \ln2$ of work, as demonstrated theoretically~\cite{Landauer_1961} and experimentally~\cite{Berut2012, Berut2015, orl12, Bech2014, Gavrilov_EPL_2016, Finite_time_2020, Hong_nano_2016, mar16, DAGO}, with $k_B$ Boltzmann's constant. Practical implementations require an overhead to Landauer's Bound (LB), observed to scale as $k_B T_0 \times B/\tau$, with $\tau$ the protocol duration and $B$ close to the system relaxation time~\cite{Finite_time_2020}. Most experiments use overdamped systems, for which minimizing the overhead means minimizing the dissipation. Underdamped systems \cite{DAGO, gieseler_levitated_2018, gieseler_non-equilibrium_2015} therefore seem natural to reduce this energetic cost. But cutting the dissipative energy cost has a counter part: in this letter we show experimentally and theoretically that, for such systems, fast erasures induce a heating of the memory and an accordingly higher energy expense, $k_B\Teff \ln{2}$. The work influx is indeed not instantaneously compensated by the inefficient heat transfer to the thermostat, which results in a transient temperature rise visible in the kinetic and potential energy evolutions. Our model covering all damping regimes paves the way to new optimisation strategies \cite{deshpande}, based on the thorough understanding of the energy exchanges.
 
\begin{figure}[ht]
\includegraphics[width=\columnwidth]{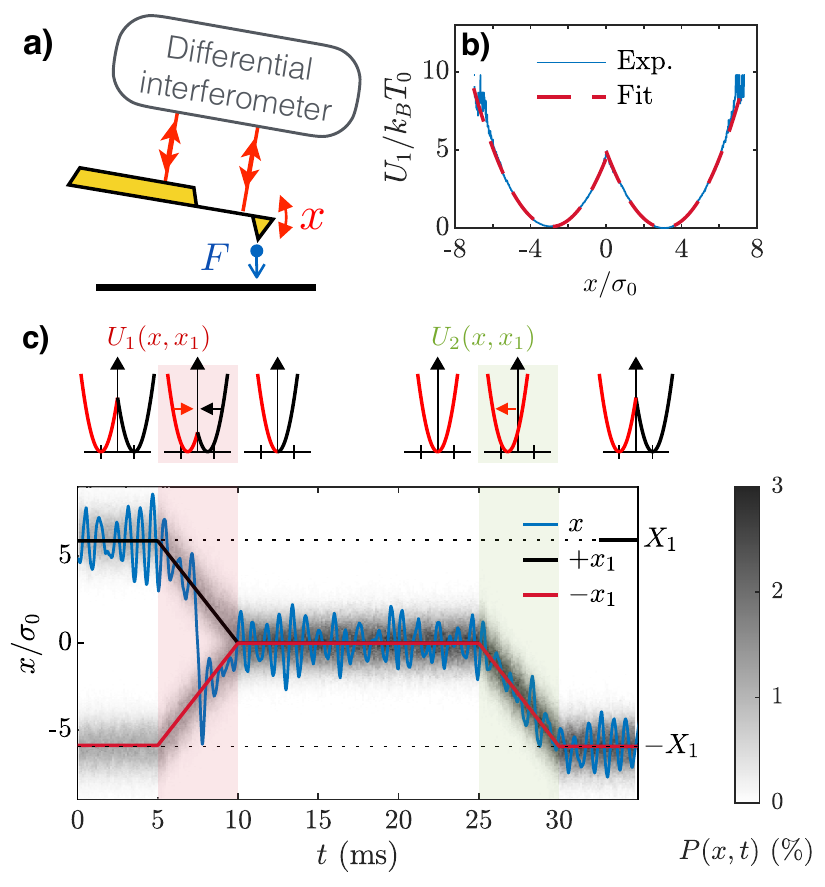}
\caption{\textbf{Experimental set up and fast erasure cycle}
(a) Schematic diagram of the experiment: the underdamped oscillator (resonance frequency $f_0=\omega_0/(2\pi) =\SI{1270}{Hz}$, quality factor $Q=10$) is a conductive cantilever, sketched in yellow. Its deflection $x$ is measured with a differential interferometer~\cite{Paolino2013}. The potential $U$ is created by the electrostatic force $F$ between the cantilever and the electrode facing it~\cite{DAGO}. 
(b) Measured double-well potential energy $U_1$ (blue) when $x_1=\sqrt{10}\sigma_0$ ($5\,k_BT_0$ barrier height), with $T_0=\SI{295}{K}$ and $\sigma_0 = \sqrt{k_BT_0/k} \sim \SI{1}{nm}$ the standard deviation of the deflection at equilibrium inside a single well. The potential is inferred from the measured PDF of $x$ during a $\SI{10}{s}$ acquisition and the Boltzmann distribution. The fit to the $U_1(x,x_1)$ expression is excellent (dashed line).
(c) Time recording of the cantilever deflection $x$ on a single trajectory (blue, starting in state 1 in this example), superposed with the centers of two wells: $+x_1$ (black) and $-x_1$ (red). Snippets of the potential energy on top of the plot sketch the erasure protocol. Stage 1 (red background) et 2 (green background) both last $\tau=\SI{5}{ms}$. The equilibrium periods around stages 1 and 2 are chosen freely as long as they allow the cantilever to relax (natural relaxation time: $\tau_r=2Q/\omega_0 \sim \SI{2.5}{ms}$). The grey map corresponds to the PDF of x, computed from $N=2000$ experimental trajectories of the erasure process.} \label{cycle_rapide}
\end{figure}

The system under scrutiny is illustrated in Fig.~\ref{cycle_rapide}: an underdamped micro-mechanical oscillator confined in a double-well potential $U_1(x,x_1)= \frac{1}{2}k(| x | -x_1)^2$, with $x$ the position of the oscillator, $k$ its stiffness and $x_1$ the user-controlled parameter tuning the barrier height~\cite{DAGO}. In our study, the only relevant degree of freedom of the physical system, a micro-cantilever, is its first deflection mode. The 1-bit information is encoded in the mean position: using a large barrier ($x_1 = X_1 \gg \sigma_0 = \sqrt{k_B T_0/k}$), the system can be at equilibrium either in the state 0 (in the left-hand well centered in $-X_1$) or in the state 1 (in the right-hand well centered in $+X_1$). The erasure process (illustrated in Fig.~\ref{schemaprot} in the Suppl. Mat.) consists in lowering the barrier and merging the wells (stage 1: decreasing $x_1$ from $X_1$ to $0$ in a time $\tau$), then translating the single well $U_2(x,x_1)= \frac{1}{2}k(x + x_1)^2$ to position $-X_1$ (stage 2: increasing $x_1$ from $0$ to $X_1$ in a time $\tau$), before recreating the second well centered in $+X_1$ to recover the initial potential $U_1$~\cite{DAGO}. The experimental probability distribution function (PDF) evolution in grey on Fig.~\ref{cycle_rapide}c points out the 100\% success rate: this protocol always drives the system in state 0 independently of its initial state.

Along a trajectory, the total energy of the system consists in the sum of the potential energy $U$ and of the kinetic energy $\K=\frac{1}{2}m v^2$ (with $m$ the oscillator mass and $v=\dot{x}$ its speed): $E=U+\K$. This quantity equilibrates with the stochastic work $\W$ and heat $\Q$ through the energy balance equation: 
\begin{align}
\frac{dE}{dt}=\frac{dU}{dt}+\frac{dK}{dt}=\frac{d\W}{dt}-\frac{d\Q}{dt}. \label{balance}
\end{align}
This energy balance is the starting point of the model developed in this Letter to explain the heating of the memory during erasure, similarly to the approach followed in the theoretical description of feedback cooling~\cite{gieseler_levitated_2018,gieseler_non-equilibrium_2015}.

The data plotted in Fig.~\ref{cycle_rapide}c ($x$ and $x_1$ along a trajectory) contains all we need  to compute the quantities involved in Eq.~\eqref{balance}. Indeed, applying to the underdamped regime the generic computations of stochastic energy exchanges~\cite{DAGO,sek10,sek66,Aurell_2012,Seifert_2012,Jarzynski_2011,Ciliberto_PRX}, we have: 
\begin{align}
\frac{d\W}{dt}&=\frac{\partial U}{\partial x_1} \dot{x}_1, \label{dWsdt} \\ 
\frac{d\Q}{dt}&=-\frac{\partial U}{\partial x} \dot{x} -\frac{d\K}{dt}. \label{dQsdt}
\end{align}

\begin{figure}[ht]
\includegraphics[width=\columnwidth]{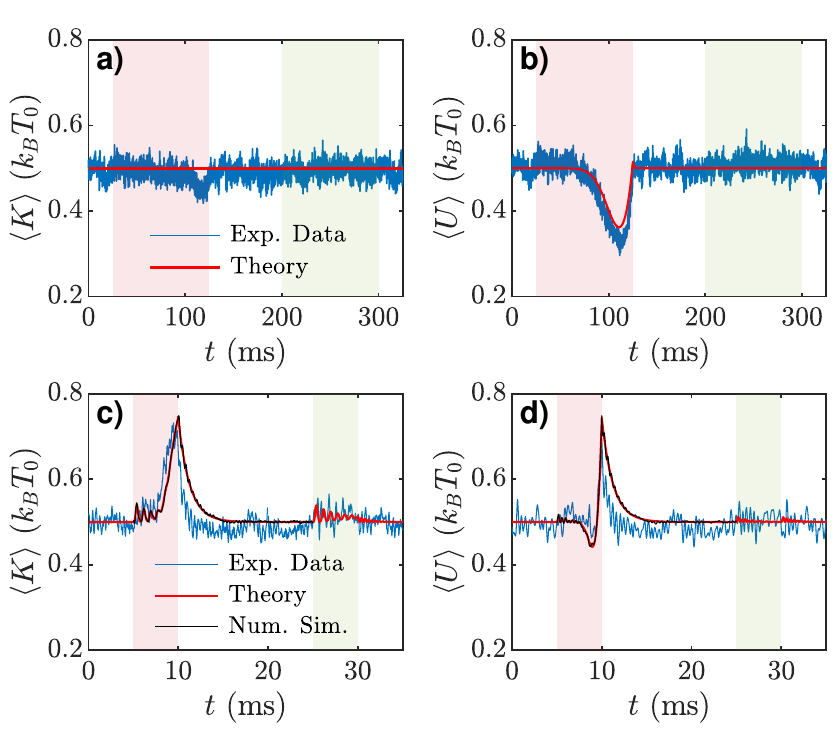}
\caption{\textbf{Energy evolution during an erasure procedure}
(a) In blue, the time evolution of the mean kinetic energy $\langle \K \rangle $ in $k_B T_0$ units over $N=2000$ iterations of a quasi-static erasure ($\tau f_0 \gg 1$): stage 1 and stage 2 (red and green backgrounds) both last $\tau=\SI{100}{ms}$. The red line corresponds to the theoretical prediction detailed in the Letter.
(b) Same, with the mean potential energy $\langle U \rangle$.
(c) and (d) Same for a fast erasure: $\tau=\SI{5}{ms}$. We add in black line the results of a numerical simulation for step 1 that provides more samples than the experiment, $N_\mathrm{sim}=\num{5e6}$ and is thus free of statistical uncertainty.} \label{Energie rapide}
\end{figure}

In Ref.~\onlinecite{DAGO} we measure the mean work and dissipated heat of erasure processes at different ramp speeds, starting from the quasi-static regime ($\tau f_0=250$), to very fast erasures ($\tau f_0\sim6$ in Fig.~\ref{cycle_rapide}c). As stated in the introduction, reducing the operation time requires an overhead to LB: the mean work and heat on an erasure cycle are, to a first approximation, $\langle \W \rangle = \langle \Q \rangle \sim k_BT_0 (\ln 2 +B/\tau)$. In this Letter, we explain the origin of this overhead increasing with the speed: it comes in underdamped memories from the transient rise of the effective temperature $\Teff$, a source of energy loss that fundamentally differs from the viscous dissipation contribution of overdamped systems.

For this purpose, we measure the mean kinetic and potential energy during either a quasi-static erasure (Fig.~\ref{Energie rapide}a-b) and a fast one (Fig.~\ref{Energie rapide}c-d). When we proceed in a quasi-static fashion, the mean kinetic energy stays as expected at its equilibrium value $\frac{1}{2}k_B T_0$, while odd evolution of the mean potential energy complies with equipartition for the bi-quadratic shape of $U_1$ as detailed in section \ref{suppmatEP} of the Suppl. Mat. For fast operations, the energy profiles are completely different: in particular they strongly increase during stage 1, before relaxing during the equilibration step. $\K$ can be decomposed into $\langle \K \rangle = \frac{1}{2} m (\langle v \rangle ^2 + \sigma_v^2$), summing the contribution of the velocity mean value $\langle v \rangle $ that reflects the response to the well motion, and the velocity variance $\sigma_v^2$ which defines the kinetic temperature (following Ref.~\cite{martinez_adiabatic_2015}) of the first deflection mode: $T = m \sigma_v^2/k_B$ . The first term is responsible for the transient oscillations at the beginning of step 1 and during step 2, but the energy rise during step 1 mainly comes from the thermal term: $\frac{1}{2} m \langle v \rangle ^2 \sim \frac{1}{2} k_B T_0 X_1^2/ (2 \pi \sigma_0 \tau f_0)^2 \ll \frac{1}{2} k_B T_0$. It therefore demonstrates a transient temperature rise.

This warming and its consequences on the operation cost can be interpreted using a simple analogy: during step 1, the system behaves as a single-particle gas~\cite{Bennett-1987} at pressure $p$, compressed so that the available volume $\V$ is divided by 2. The infinitesimal work required for the compression is $d\W^c=-p d\V=- k_B T d\ln \V$. If the transformation is quasi-static, $T=T_0$ and the work simplifies into $\W^c=k_B T_0\ln2$. On the contrary, if the process is too fast to allow heat exchanges with the surrounding thermostat, the transformation is adiabatic, and the temperature $T$ of the gas increases during the compression. Hence, the compression work for fast operations writes $\W^c=k_B \int_0^\tau T d\ln \V = k_B \Teff \ln2$ with $\Teff \geq T_0$. The heat exchanges after the adiabatic compression will then allow the system to thermalize at $T_0$.

By analogy, we will also call "compression" the reduction of the phase space volume explored when the bi-stable potential progressively shrinks until reaching a single well during step 1. This analogy highlights the fact that the warming during the compression is specific to the underdamped system, and would not exist if a strong coupling to the bath allowed efficient heat exchanges. The objective of the following sections is to build a model which describes both the compression and translational motion as observed in experiment.

We first proceed with the mean dissipated heat, described by (details in section \ref{suppmatHeat} of the Supp. Mat.):
\begin{align}
\frac{d\langle \Q\rangle}{dt}&=\frac{\omega_0}{Q}\big(m\langle v \rangle^2  + k_B (T - T_0)\big). \label{Qgen}
\end{align}
This expression is completely general and highlights that the heat exchanges are reduced at high quality factors $Q$.

To compute the other energetic terms ($\langle \W \rangle$, $\langle K \rangle$ and $\langle U \rangle$), we rely on the PDF of position $x$ and speed $v$. Let us introduce this PDF during the compression stage, supposing that the system is at equilibrium: it is governed by the Boltzmann distribution
\begin{subequations} \label{eq.PZc}
\begin{align}
P^{c}(x,v) &= \frac{1}{Z^{c}} e^{-\frac{1}{2} \beta m v ^2} e^{-\frac{1}{2} \beta k (| x | -x_1)^2} \label{Pc}\\
Z^{c}(\beta,x_1) &= \frac{2 \pi}{\sqrt{km} \beta}V, \quad V=1+\erf\left(\sqrt{\frac{k\beta}{2}} x_1\right),  \label{Zc}
\end{align}\end{subequations}
with $\beta=1/ (k_B T)$, $Z^{c}$ the partition function, and $V$ a volume-like function that shrinks by a factor 2 when $x_1$ decreases from $X_1$ to $0$. We can directly apply this PDF to the slow erasures, in equilibrium at temperature $T_0$ at all time. We extend the use of this PDF to the case of fast erasures as well, under the hypotheses that (i) the cantilever oscillates several times in the double-well before its shape changes significantly ($|\dot x_1| \ll \omega_0\sigma_0$), so that the phase space is adequately sampled and (ii) a Boltzmann-like distribution still holds. In this case, however, we let the temperature $T$ as a parameter free to evolve due to a possible heating. Note that the PDF $P^c(x,v)$ only describes the volume compression and does not include any transients, leaving aside any coupling between $x$ and $v$. The main transient, due to the translational motion of the wells, is addressed in the next paragraph. In section \ref{suppmatPDFansatz} of the Suppl. Mat., we compare the PDF of our ansatz with one sampled on a large numerical simulation, demonstrating its relevancy.

During stage 2, or at the beginning of stage 1 before the oscillator crosses the barrier, the dynamics is ruled by a linear Langevin equation: the potential energy is quadratic (no switching). $x(t)$ is therefore the sum of the stochastic response to the thermal fluctuations, and of the deterministic response $\pm x_D(t)$ to the driving force $F_D(t)=\pm k x_1(t)$ (the sign depending of which well is considered). $x_D(t)$ can be easily computed for our simple $x_1(t)$ ramps, and the PDF $P^t(x,v)$ which determines the translational motion is then described by~\cite{LeCunuder,martinez_supplemental_2015}:
\begin{subequations}\label{eq.PZt}\begin{align}
P^t(x,v) &= \frac{1}{Z^t} e^{-\frac{1}{2} \beta m (v -\dot x_D)^2} e^{-\frac{1}{2} \beta k (x -x_D)^2}   \label{P1t}\\
Z^t &= \frac{2 \pi}{\sqrt{km} \beta} V, \quad V=1.
\end{align}\end{subequations}
We easily retrieve $\langle x \rangle=x_D$ and $\langle v \rangle=\dot x_D$.

In complement to Eq.~\eqref{Qgen} for the mean heat, the knowledge of the PDF allows the computation of all mean energetic quantities. During compression for example, the mean energy is $\langle E^c \rangle = -\partial \ln Z^c / \partial \beta $, while the mean work derivative is $\langle \dot \W^c \rangle = \langle \partial U /\partial x_1 \rangle \dot x_1 = -\dot x_1 / \beta \, \partial \ln Z^c / \partial x_1 $. In section \ref{suppmatE} of the Suppl. Mat., we derive the following expressions, valid for all stages:
\begin{subequations} \label{eq.E}
\begin{align}
\frac{d\langle \Q \rangle}{dt}=& \frac{\omega_0}{Q}\big(2 K_D  + k_B T - k_B T_0\big) \\
\frac{d\langle \W \rangle}{dt} =& \frac{d \W_D}{dt} -  k_B T \frac{\partial \ln \V}{\partial x_1} \dot{x}_1\label{eq.W}\\
\langle K \rangle =& K_D + \frac{1}{2} k_B T \label{eq.K}\\
\langle U \rangle =& U_D + \frac{1}{2} k_B T  + k_B T^2 \frac{\partial \ln \V}{\partial T} \label{eq.U}
\end{align}
\end{subequations}
where $\W_D$, $K_D$ and $U_D$ are respectively the deterministic work, kinetic and potential energy which vanish in the quasi-static regime. With Eq.~\eqref{eq.W} for a quasi-static compression in equilibrium at $T_0$, we recover the gas analogy $d\W^c=- k_B T_0 d\ln \V$, hence LB:  $\langle \W^{c} \rangle =k_B T_0 \ln 2$. 

Using Eqs.\eqref{balance} and \eqref{eq.E}, we derive a differential equation governing the time evolution of the temperature: the deterministic terms cancels out, since they comply to the energy balance as well, and we're left with 
\begin{equation}
\begin{split}
\frac{d\langle  E  \rangle}{dt} &= \frac{\partial \langle E  \rangle}{\partial T}\dot T + \frac{\partial \langle E \rangle}{\partial x_1}\dot x_1 \\
&= - k_B T \frac{\partial \ln \V}{\partial x_1} \dot{x}_1+\frac{k_B \omega_0}{Q}(T-T_0).
\end{split}
\end{equation}

Explicit formulas for $\partial \langle E  \rangle / \partial T$ and $\partial \langle E  \rangle/\partial x_1$ are readily computed from Eqs.~(\ref{eq.K}-\ref{eq.U}). When we proceed in quasistatic fashion ($\dot{x}_1\sim 0$), or when the volume is constant ($\partial /\partial x_1=0$) , we observe no heating: $T=T_0$. For fast compressions, this equation can be solved numerically and leads to the evolution of the kinetic temperature $T(t)$.

\begin{figure}[t]
\includegraphics[width=\columnwidth]{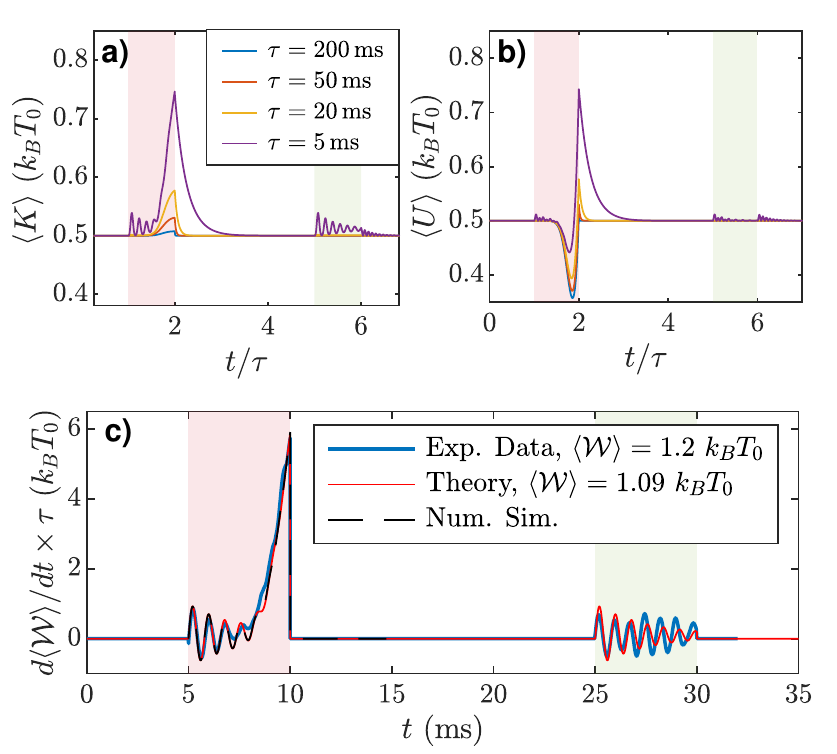}
\caption{\textbf{Model prediction: energy and  stochastic work profiles} (a) Time evolution of the mean kinetic energy $\langle \K \rangle$ for different duration $\tau$ of the erasure steps computed from Eq.~\eqref{eq.K}. For small $\tau$, $\langle \K \rangle$ is affected during step 1 (red background) by a transient oscillation due to the dragging, followed by a strong rise in temperature. Only the dragging transient appears during step 2 (green background). (b) Same plot for the potential energy $\langle U \rangle$ from Eq.~\eqref{eq.U}. (c) Time evolution of the mean power over 2000 trajectories, following the fast protocol ($\tau=\SI{5}{ms}$) corresponding to Fig.~\ref{cycle_rapide} (blue). The red line is computed using Eq.~\eqref{eq.W} and closely matches the experimental results. Results of a numerical simulation (black dashed line), corresponding to $\num{5e6}$ trajectories, match the model so well that we cannot distinguish the curves.}
\label{Energieth}
\end{figure}

Thanks to the knowledge of $T(t)$, our model describes the evolution of all energetic quantities in Eqs.~\eqref{eq.E} during the erasure process. For slow erasures, kinetic (Fig.~\ref{Energieth}a) and potential (Fig.~\ref{Energieth}b) energies comply as expected with equipartition. For fast erasures, we obtain a strong temperature increase~\footnote{Since $< v >^2 \ll \sigma_v$, $< K > \sim \frac{1}{2} k_B T$ and the temperature profile can be read directly on the kinetic energy curve.} during step 1, visible on both energy profiles. The system then thermalizes, before responding to the translational motion of step 2 with transient oscillations. Those theoretical results superimposed on Fig.~\ref{Energie rapide} in red lines are in very good agreement with the experimental observations for both slow and fast erasures, with no adjustable parameters. We supplemented the model validation by numerical simulation data (see Suppl. Mat. section \ref{suppmatnumsim}): the black curve on Fig.~\ref{Energie rapide}c-d closely matches the model, except for tiny ripples during the thermalization that correspond to transients unaccounted for. Additionally, the model predicts that a fast erasure cycle will cause a mean power evolution that displays transient oscillations and a rise during compression, both of which are consistent with the experimental data of Fig.~\ref{Energieth}c, and perfectly matches the simulation results.

\begin{figure}[ht]
\includegraphics[width=\columnwidth]{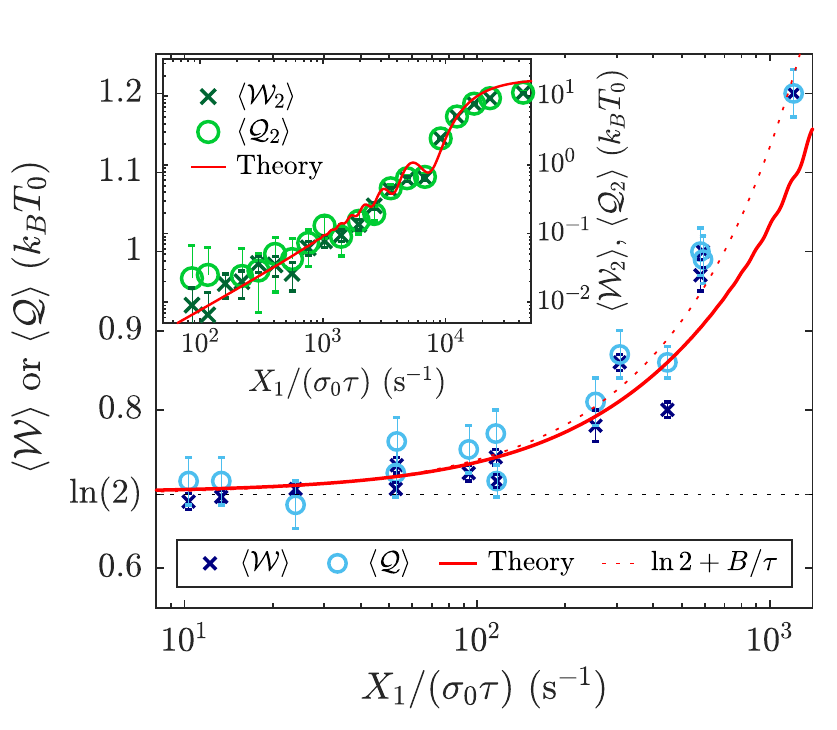}
\caption{\textbf{Divergence from the Landauer limit for fast erasures.} Erasure cost  ($\langle \W \rangle $ and $\langle \Q \rangle $ in $k_B T_0$ units) for different operation speeds $X_1/(\sigma_0 \tau)$. Experimental data (in blue) computed from $N=2000$ iterations each with $X_1\sim6\sigma_0$, are in good agreement with the analytical computation (red line). As a comparison, we plot the empirical description of the divergence from LB as $k_BT_0(\ln{2}+B/\tau)$ used in the existing literature, with $B=(2.6\pm0.2)\SI{}{ms}$ here (red dotted line). Inset: same considering only the translational motion in step 2.}
\label{B_tau}
\end{figure}

All in all, we propose an efficient theoretical framework to predict the energy exchanges and explore the fast information erasure cost. The model only requires the system parameters ($f_0$ and $Q$) and the protocol ones ($X_1$ and $\tau$) to estimate the erasure cost. As a further illustration of the model reliability, in Fig.~\ref{B_tau} we compare its predictions with the experimental points and the empirical description of the overhead in $B/\tau$: it successfully quantifies the  divergence from LB as the speed is increased. The remaining difference may result from calibration drifts or experimental imperfections~\cite{Dago-Deamon}, or from the shortcomings of the model with respect to transients.

Furthermore, the model distinguishes the part of the overhead due to the compression to the one due to the translational motion. The latter, plotted in the inset of Fig.~\ref{B_tau}, behaves at first order as $X_1^2/(Q\omega_0)\tau$s~\cite{DAGO,Finite_time_2020}. On the contrary, the former increases with the quality factor $Q$: it behaves as $k_B\Teff \ln 2$, with $\Teff>T_0$ the effective temperature during the process (details in section \ref{suppmatTeff} of the Supp. Mat.), rising when the heat exchanges with the bath are reduced (at high speeds or high $Q$). In the mean adiabatic limit (as defined in Ref.~\onlinecite{martinez_adiabatic_2015}) for erasure, however, the compression work saturates at $k_BT_0$, as shown section \ref{suppmatadiabiatic} of the Supp. Mat. Indeed, for adiabatic transformations of underdamped systems, the conservation of the phase space volume~\cite{martinez_adiabatic_2015} requires to enslave the variations of the temperature $T$ to those of the volume $V$. These considerations open several possibilities (that could be combined) to optimise the information processing: applying optimal protocols for the translational motion (predominant for overdamped systems) as suggested in Refs. \onlinecite{Gomez,Dellago,Finite_time_2020}; moving to the underdamped regime to reduce the operation time scale and decrease the dragging cost~\cite{DAGO}, while paying only the adiabatic limit $k_B T_0$ for $Q\gg1$. 

As a conclusion, the underdamped framework addressed in this Letter opens up new possibilities in information processing: the operation times are several orders of magnitude smaller than the ones encountered in the overdamped regime, as well is the cost required to move the system in the bath. Nevertheless, the price to pay to get rid of the viscous slowdown hides in the low coupling to the bath, allowing the memory temperature to strongly rise for fast drivings. We provide a full theoretical description of an erasure cycle which results are verified by a wide panel of high-resolution experimental measurements and complementary numerical simulations. It culminates in the prediction of the overhead to LB for fast erasures. Such an understanding of the erasure process, covering all damping regimes, paves the way to new approaches to the information processing optimisation.

\acknowledgments

The data that support the findings of this study are openly available in Zenodo~\cite{Dago-2021-DatasetPRL,Dago-2021-Dataset2}.

\medskip
This work has been financially supported by the Agence Nationale de la Recherche through grant ANR-18-CE30-0013. We thank S. Ciliberto, J. Pereda and N. Barros for fruitful scientific discussions and advices.

\bibliographystyle{apsrev4-2}
\bibliography{DynInfoErasure}

\clearpage

\onecolumngrid

\renewcommand{\thepage}{S\arabic{page}}
\renewcommand{\thesection}{S\arabic{section}}
\renewcommand{\thetable}{S\arabic{table}}
\renewcommand{\thefigure}{S\arabic{figure}}
\renewcommand{\theequation}{S\arabic{equation}}
\setcounter{page}{1}
\setcounter{section}{0}
\stepcounter{AppendixEquation}
\stepcounter{AppendixFigure}
\thispagestyle{empty}

\begin{center}
{\large \bf \textsc{Supplemental material}\\Dynamics of information erasure and extension of Landauer's bound to fast processes}

\vspace{4mm}

Salamb\^{o} Dago and Ludovic Bellon

{\it \small Univ Lyon, ENS de Lyon, CNRS, Laboratoire de Physique, F-69342 Lyon, France}
\end{center}

\vspace{4mm}

\twocolumngrid

\stepcounter{AppendixEquation}
\stepcounter{AppendixFigure}

\section{Experiment}

The experimental setup is described in Refs.~\onlinecite{DAGO, Dago-Deamon}, and we summarise in this paragraph the main characteristics. The underdamped oscillator is a conductive cantilever (Doped silicon cantilever OCTO 1000S from Micromotive Mikrotechnik, $\SI{1}{mm}$ long, $\SI{90}{\mu m}$ large, $\SI{1}{\mu m}$ thick) in air at room temperature. Its deflection is measured with a differential interferometer~\cite{Paolino2013}, which features a high stability and high resolution ($\sim\SI{1}{pm}$ for a $\SI{10}{kHz}$ bandwidth). To modify the energy potential seen by the oscillator, we create an electrostatic force by applying a voltage $V$ to a nearby electrode. The value of $V$ is set by a feedback loop on the position $x$ measured by the interferometer~\cite{Bech2014,DAGO}: $V=V_0+S(x-x_0)V_1$, where $S$ is the sign function. The threshold $x_0$ is chosen midway between the two equilibrium positions to create the symmetric double-well, or chosen far above (or below) to force one single well. The amplitude of $V_1$ tunes the distance between the wells, thus $x_1$ in our protocols. The sign function is implemented with a high speed analog comparator (response time below $\SI{1}{\mu s}$). To avoid multiple fast switching of the comparator around the threshold because of measurement noise, we apply a short temporal lockup: once triggered, the comparator cannot switch back to the preceding state during the next quarter period $f_0^{-1}/4\sim\SI{200}{\mu s}$. During this lockup period, the oscillator is expected to have reached a position far enough from the threshold before rearming the trigger. This strategy allows the feedback potential experienced by the oscillator to be equivalent to the prescribed double-well~\cite{DAGO, Dago-Deamon}.

\section{Numerical simulation} \label{suppmatnumsim}

The experimental results are supplemented by a numerical simulation providing a large number of trajectories ($N^\mathrm{sim}=\num{5e6}$) without any calibration drift in the initial position $X_1$. The simulation is meant to mimic the experimental system during step 1, and therefore uses the experimental parameters $\omega_0$, $m$, $Q$, $X_1$ and $\tau$. The simulation code consists in integrating the Langevin equation that rules the cantilever position:
\begin{align}
\ddot x +\frac{\omega_0}{Q} \dot x + \omega_0^2 x&=\frac{F_{th}}{m}+S(x) \omega_0^2  (X_1-\frac{X_1}{\tau}t ), \label{Langevin}
\end{align}
where $F_{th}$ the stochastic forcing from the bath, satisfying $\langle F_{th}(t)  F_{th}(0)\rangle = \delta(t) 2 k_B T_0 m\omega_0 /Q$. $F_{th}$ is implemented as a random number normally distributed around $0$, with a standard deviation $\sqrt{2 k_B T_0 m\omega_0 /(Q\Delta t)}$, with $\Delta t$ the simulation time step. 
We choose the symplectic Euler method~\cite{Eulersymp}, better suited to stochastic differential equation than the Runge-Kutta one~\cite{PhysRevE.74.068701}, to solve numerically Eq.~\eqref{Langevin} and output the position and speed of the cantilever at every time step.  We display in Listing~\ref{code} the first steps of the symplectic Euler method with 
normalized quantities, $z=x/\sigma_0$ and $s=\omega_0t$:

\begin{lstlisting}[language=Python, caption=Symplectic Euler method,label=code]
Fth=sqrt(2/(Q*ds))*randn(N)
for i in range(N-1)
	z[i+1]=z[i]+v[i]*ds
	dz=z[i+1]-sign(z[i+1])*z1[i+1]
	v[i+1]=v[i]-(dz+v[i]/Q+Fth[i])*ds
\end{lstlisting}

The initial position and speed are distributed according to the Boltzmann equilibrium distribution, corresponding to Eq.~\eqref{eq.PZc} with $\beta=1/{k_B T_0}$ and $x_1=X_1$. 

\begin{figure*}[htb]
 \includegraphics[width=15cm]{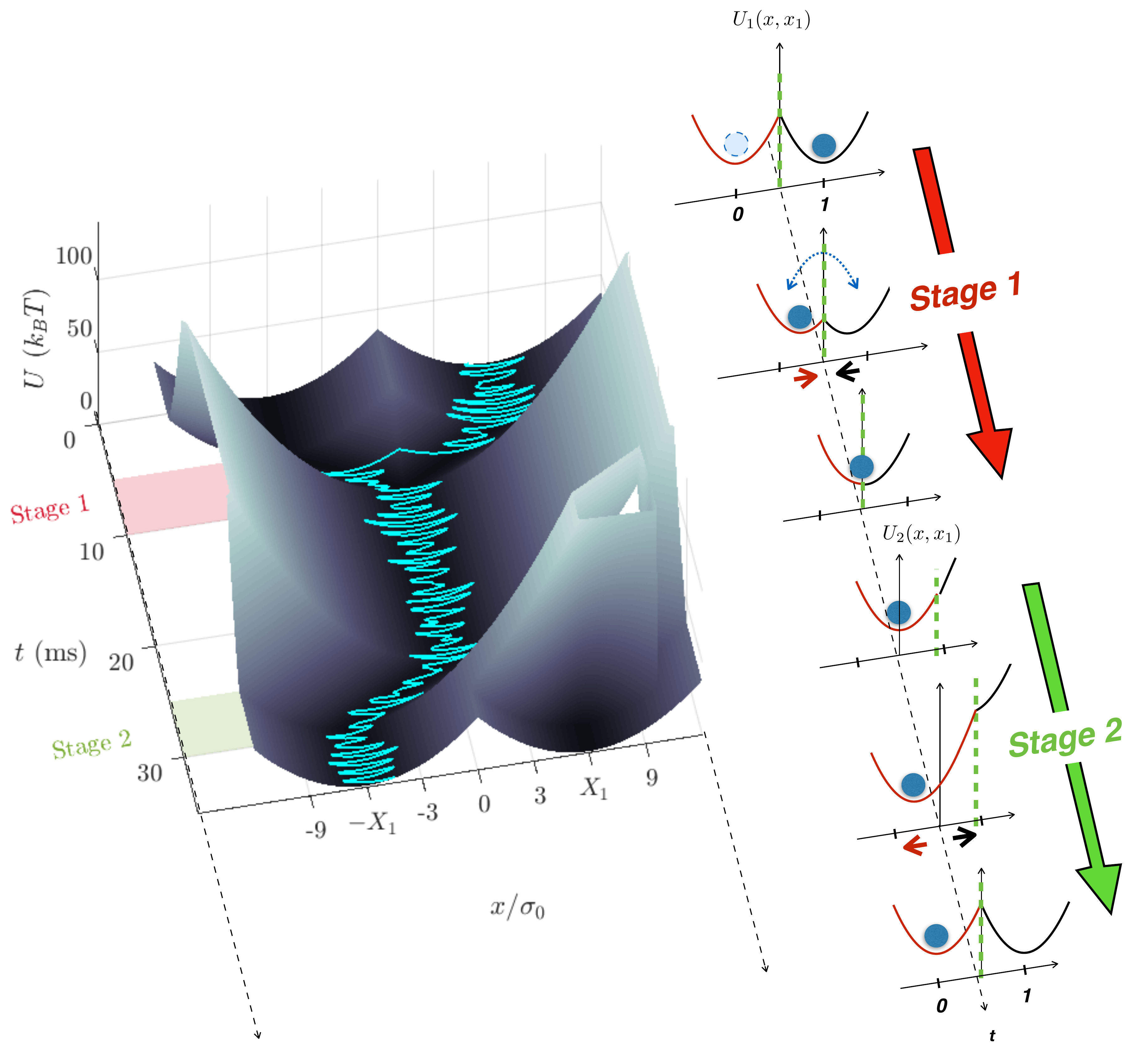}
 \caption{{\bf Erasure protocol} The two wells initially at a distance $2X_1$, high enough to secure the bit initial value, are brought together during stage 1 until they merge into one single well in $x=0$. Then, during stage 2, the memory is reset by bringing back the well center to $-X_1$, before rebuilding the bi-stable potential with the barrier at position $x=0$. A trajectory is superimposed on the energy surface in a 3D spatio-temporal representation on the left, while snippets of the potential energy are shown on the right to sketch the protocol.}
\label{schemaprot}
\end{figure*}

\section{Erasure protocol}

The erasure protocol is the one described in Ref.~\onlinecite{DAGO}:  stage 1 consists in lowering the barrier and merging the wells (decreasing $x_1$ from $X_1$ to $0$ in a time $\tau$), then stage 2 consists in translating the single well to position $-X_1$ ( increasing $x_1$ from $0$ to $X_1$ in a time $\tau$), before recreating the second well centered in $+X_1$ to recover the initial potential $U_1$. Fig.~\ref{schemaprot} maps the protocol evolution during the erasure process, with $U_1(x,x_1)= \frac{1}{2}k(| x | -x_1)^2$ and $U_2(x,x_1)= \frac{1}{2}k(x + x_1)^2$  respectively the stage 1 and stage 2 potential expression.

\section{Energetic quantities computation} \label{suppmatE}
\subsection{Equipartition theorem in a double well potential} \label{suppmatEP}

We apply in this section the equipartition theorem to the underdamped cantilever evolving in a double well potential $U_1(x,x_1)= \frac{1}{2}k(| x | -x_1)^2$. The total energy of the system $E=U_1+K$ (with $K=\frac{1}{2}mv^2$) is given at thermal equilibrium (at bath temperature $T_0$) by the very general equipartition formula:
\begin{align}
\langle x_m \frac{\partial E}{\partial x_n}\rangle=\delta_{nm}k_BT_0,
\label{equipart}
\end{align}
where the $x_n$ are the degrees of freedom of the system.

In our 1D description of the underdamped cantilever, we consider only two degrees of freedom:  the velocity $v=\dot x$ and the position $x$. The equipartition theorem \eqref{equipart} applied to the velocity degree of freedom results in:
\begin{align}
\langle v \frac{\partial E}{\partial v} \rangle = \langle v \frac{\partial K }{\partial v} \rangle= 2 \langle K \rangle = k_BT_0
\label{Ec}
\end{align}
For the second degree of freedom, the position $x$, we derive:
 \begin{align}
\langle x \frac{\partial E}{\partial x} \rangle = \langle x \frac{\partial U_1}{\partial x} \rangle = k \langle x^2-|x |x_1 \rangle&=k_BT_0\\
2 \langle U_1 \rangle+ kx_1(\langle |x| \rangle -x_1)&=k_BT_0
\label{Ueqpart}
\end{align}
The term $\langle |x| \rangle$ can be computed using the equilibrium probability distribution, as detailed in the Supplementary Materials of \cite{DAGO} (introducing $\sigma_0=\sqrt{k_BT_0/k}$):
\begin{align}
\langle |x|  \rangle=\sigma_0\sqrt{\frac{2}{\pi}}\frac{e^{-\frac{x_1^2}{2\sigma_0^2}}}{(1+\erf{\frac{x_1}{\sqrt{2}\sigma_0}})}+x_1
\end{align}
Finally, pluging in the above expression into Eq.~\eqref{Ueqpart} gives the mean potential energy:
 \begin{align}
  \langle U_1 \rangle=k_BT_0(\frac{1}{2} -\frac{x_1}{\sigma_0}\frac{e^{-\frac{x_1^2}{2\sigma_0^2}}}{\sqrt{2\pi}(1+\erf{\frac{x_1}{\sqrt{2}\sigma_0}})})
  \label{meanU}
\end{align}
 
Eq.~\eqref{meanU} deduced from the equipartition theorem explains the dip in the potential energy in the quasi-static regime when $x_1$ decreases during stage 1.
 Eq.~\eqref{meanU} is not intuitive because we are used to the equipartition result in the harmonic potential case: $\langle U \rangle =\frac{1}{2}k_BT_0$. Nevertheless for a bistable potential the equipartition theorem leads to a more complex expression. Let us point out here that for $x_1=0$ and $x_1=+\infty$, Eq.~\eqref{meanU} tends to the harmonic result.
 
\subsection{Mean heat} \label{suppmatHeat}
The computation of the mean dissipated heat requires writing the general Langevin equation of an underdamped system in a potential $U$:
\begin{equation}
m\ddot x = -\frac{\partial U}{\partial x} -\frac{m\omega_0}{Q} \dot x  +F_{th} \label{langevingen}
\end{equation}
Multiplying Eq.\eqref{langevingen} by $\dot x$ leads to the dissipated heat defined by Eq.~\eqref{dQsdt}:
\begin{align}
\frac{d\Q}{dt}=m\ddot x \dot x -\frac{d\K}{dt} +\frac{m\omega_0}{Q} \dot x ^2  -F_{th}\dot x \label{computeQ}
\end{align}

Some caution is required before taking the mean value of the above expression, because it involves products of stochastic quantities: in that respect, the Ito discretization prescribes for a stochastic function $f(v)$, 

\begin{align}
\frac{df}{dt}=\frac{\partial f}{\partial v} \dot v +\frac{1}{2} \frac{\partial^2 f}{\partial v^2}\dot v^2 dt \label{Ito}
\end{align}
If we combine Eq.~\eqref{langevingen} and  Eq.~\eqref{Ito} applied to $K=1/2 m v^2$, before taking the mean value and making $dt$ tend to $0$, most terms simplify out. Indeed, only remain the terms that involve the thermal noise scaling in $1/\sqrt{t}$, some of which are cancelled by the Ito prescription: $\langle F_{th}  v \rangle = \langle F_{th}  x \rangle=0$.  
Finally, we obtain the relation: $d \langle \K \rangle/dt = m\langle \ddot x \dot x \rangle + k_B T_0 \omega_0/Q$. Eq.~\eqref{computeQ} then simplifies into Eq.~\eqref{Qgen}.

\subsection{Mean work, potential and kinetic energy}

Let us split the PDF $P^c(x,v)$ describing compression and its associated partition function $Z^c$ of Eqs.~\eqref{eq.PZc} into their kinetic and potential contribution:
\begin{subequations}\begin{align}
P^c_K(v)=\frac{e^{-\beta K^c}}{Z^c_K}, \ &  Z^c_K=\int_{-\infty}^{+\infty} e^{-\beta K^c} dv = \sqrt{\frac{2 \pi}{m\beta}} \\
P^c_U(v)=\frac{e^{-\beta U^c}}{Z^c_U}, \ & Z^c_U=\int_{-\infty}^{+\infty} e^{-\beta U^c} dx = \sqrt{\frac{2 \pi}{k\beta}}V.
\end{align}\end{subequations}
We now easily compute the mean values of $K$ and $U$:
\begin{subequations}\begin{align}
\langle K^c \rangle &=  \int_{-\infty}^{+\infty} K^c \frac{e^{-\beta K^c}}{Z^c_K} dv = - \frac{\partial \ln Z^c_K}{\partial \beta} = \frac{1}{2\beta}, \\
\langle U^c \rangle &=  \int_{-\infty}^{+\infty} U^c \frac{e^{-\beta U^c}}{Z^c_U} dx = - \frac{\partial \ln Z^c_U}{\partial \beta} = \frac{1}{2\beta} - \frac{\partial \ln V}{\partial \beta}.
\end{align}\end{subequations}
Because there are no deterministic terms ($\langle x \rangle=0$ and $\langle v \rangle=0$) in the compression framework, those equations equivalently arise from Eqs.~(\ref{eq.K}-\ref{eq.U}).
The mean work time derivative is computed from Eq.~\ref{dWsdt}:
\begin{equation}
\langle \frac{d\W^c}{dt} \rangle = \int_{-\infty}^{+\infty} \frac{\partial U^c}{\partial x_1} \frac{e^{-\beta U^c}}{Z^c_U} dx \dot{x}_1 = - \frac{1}{\beta} \frac{\partial \ln Z^c_U}{\partial x_1} \dot{x}_1, \label{Wc}
\end{equation}
which is again equivalent to Eq.~\eqref{eq.W} in the absence of deterministic work.

The case of translation follows a similar pattern, except that we now need to include deterministic terms, since here $\langle x \rangle= x_D$ and $\langle v \rangle= \dot x_D$. We then rewrite the energies as:
\begin{align}
\langle K^t \rangle = &  \langle \frac{1}{2} m v^2 \rangle \nonumber \\
= &  \langle \frac{1}{2} m(v-\dot x_D)^2 + m \dot x_D(v- \frac{1}{2} \dot x_D) \rangle \nonumber \\
= &  \langle \frac{1}{2} m(v-\dot x_D)^2 \rangle + \frac{1}{2} m\dot x_D^2 \\
\langle U^t \rangle = &  \langle\frac{1}{2} k(x-x_1)^2 \rangle \nonumber\\
 = &  \langle\frac{1}{2} k(x-x_D)^2+\frac{1}{2} k(x_D-x_1)^2\nonumber\\
 & \qquad\qquad\qquad +k(x-x_D)(x_D-x_1)\rangle \nonumber \\
 = &  \langle\frac{1}{2} k(x-x_D)^2\rangle + \frac{1}{2} k(x_D-x_1)^2
\end{align}
The mean values of the energies are thus the sum of a deterministic and a stochastic term. The expressions of latter and of the PDF $P^t(x,v)$ in Eq.~\eqref{P1t} are those of an harmonic oscillator in the referential centered in $x_D$, which directly lead to the equipartition. Since $V=1$ during a translation, we recover the mean values anticipated by Eqs.~(\ref{eq.K}-\ref{eq.U}). The mean work time derivative is again computed from Eq.~\ref{dWsdt}:
\begin{equation}
\langle \frac{d\W^t}{dt} \rangle =\langle \frac{\partial U^t}{\partial x_1} \rangle \dot{x}_1 = -k(x_D-x_1) \dot{x}_1
\end{equation}
In this case, the mean work is purely deterministic, as expected in Eq.~\eqref{eq.W} with $V=1$. The ansatz for the PDF in the compression or translation stages thus lead to the Eqs.~\eqref{eq.E} describing all the energetic terms is any situation.

\subsection{Deterministic terms}

The trajectory $x(t)$ in a moving well decomposes into the stochastic response to the thermal fluctuations, which vanishes on average, and the response to the driving force ramp which is the solution of the following deterministic equation: 
\begin{align}
\ddot x_D +\frac{\omega_0}{Q} \dot x_D + \omega_0^2 x_D&=\omega_0^2 x_1(t) \label{Langevintrans}
\end{align}
with $x_1(t)=X_1(1-t/\tau)$ decreasing from $X_1$ to 0 during step 1.
We solve the above equation of motion, introducing $\Omega=\omega_0 \sqrt{1-1/4Q^2}$,  and obtain the deterministic trajectory $\pm x_D(t)$ (the sign depending of which well is considered): 
\begin{equation}
\begin{split}
x_D(t)=x_1(t)+\frac{X_1}{\tau} \bigg[ &\frac{1}{Q\omega_0}\left(1- e^{\frac{t\omega_0}{2Q}}\cos\Omega t \right)\\
&\left. -\frac{1-2Q^2}{2Q^2\Omega}e^{\frac{t\omega_0}{2Q}}\sin\Omega t) \right] \label{xD}
\end{split}
\end{equation}

Those results can be applied for the translational motion during step 2, and at the beginning of step 1. Indeed, as long as the cantilever hasn't left its initial well, the above description holds during step 1. After the first commutation, the cantilever switches frequently between the symmetric wells so that the deterministic terms can be neglected. We therefore introduce $\Pi(t)$ the probability that the cantilever remains in its initial well until time $t$ (see next section for its expression), in order to modulate the deterministic contribution accordingly. The deterministic work, kinetic and potential energies are then given by: 
\begin{subequations} \label{eq.D}
\begin{align}
\frac{d \W_D}{dt} =&-k(x_D-x_1)\dot x_1\times \Pi (t) \\
K_D(t) =&\frac{1}{2} m \dot x_D \times \Pi (t) \\
U_D (t)=&\frac{1}{2} k(x_D-x_1)^2 \times \Pi (t)
\end{align}
\end{subequations}

During step 2, $x_D(t)$ is still described by Eq.~\eqref{xD} with $x_1(t)=-X_1 t/\tau$, and the energetic terms correspond at all time to Eqs.~\eqref{eq.D} with $\Pi(t)=1$ as the cantilever remains in the single well allowed.

\begin{figure*}[t]
 \includegraphics[width=\textwidth]{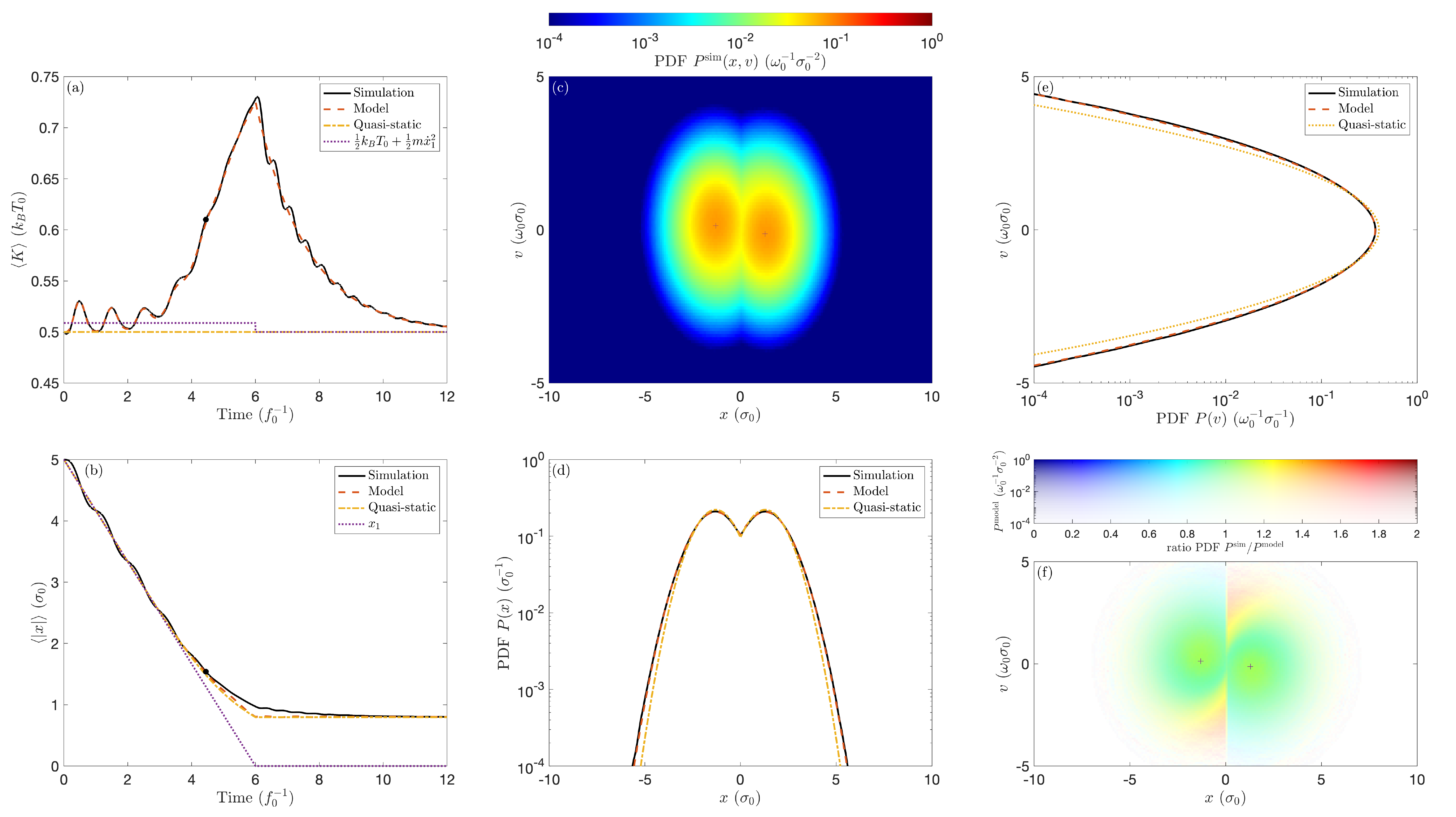}
 \caption{{\bf Comparison of the PDFs $P^\mathrm{model}(x,v,t)$ from our ansatz and $P^\mathrm{sim}(x,v,t)$ from a numerical simulation.} This frame is captured from the movie {\tt PDF\_erasure\_process.mov} (ancillary file), corresponding to $t=4.5f_0^{-1}$. (a) Mean kinetic energy $\langle K \rangle$ vs time from the simulation and the model. The quasistatic case is shown for comparison. (b) Average absolute value of the position $\langle |x| \rangle$ vs time (c) 2D representation of the PDF in $P^\mathrm{sim}(x,v,4.5f_0^{-1})$ with the color scale on top. (d) PDFs in position (obtained by integrating over all speeds the 2D PDF) from the simulation and the model. The quasistatic case is shown for comparison. (e) PDFs in speed (obtained by integrating over all positions the 2D PDF). (f) Ratio of the PDFs $P^\mathrm{sim}(x,v,4.5f_0^{-1})/P^\mathrm{model}(x,v,4.5f_0^{-1})$ in a 2D representation using the color scale on top.}
\label{Ansatz}
\vspace{5mm}

\end{figure*}

\subsection{Escape time probability distribution}

To compute $\Pi(t)$ during step 1, we tackle the escape time probability distribution of the cantilever. Indeed we want to know when on average happens the first commutation, that is to say when cantilever crosses the barrier and switches in the other well for the first time. In a good approximation, the Kramer's theory~\cite{KRAMERS} prescribes the escape rate $\Gamma (t) =\frac{\omega_0}{2 \pi} e^{-\Delta U (t)/k_BT_0}$, with the potential barrier at time $t$ being worth $\Delta U (t)= k x_1(t)^2 /2$. 
The probability that a trajectories hasn't commuted at time $t$ derives from it: 
\begin{align}
\Pi(t)&= e^{-\int_0^t \Gamma(u) du}. \label{Nt}
\end{align}
The integral in Eq.~\eqref{Nt} can be analytically expressed:
\begin{align}
\int_0^{t} \Gamma(u) du&=\frac{\omega_0\tau}{\sqrt{2\pi}} \frac{\sigma_0}{2X_1}\left[\erf(\frac{X_1}{\sqrt{2}\sigma_0})-\erf(\frac{\tau-t}{\tau}\frac{X_1}{\sqrt{2}\sigma_0})\right].
\end{align}
This estimation is consistent with the experimental data. 

\section{Validation of the PDF ansatz} \label{suppmatPDFansatz}

Using the PDFs of Eqs.~\eqref{eq.PZc} and \eqref{eq.PZt} is based on some approximations: the dragging effect is assumed to vanish after the first commutation of the system. We unavoidably leave aside some transients mixing position and speed during compression. In order to investigate on the validity of this approach we compare the numerical simulation data to the PDF models, using the following ansatz
\begin{align}
& P^\mathrm{model}(x,v,t) = \Pi(t) P^t(x,v) + [1-\Pi(t)] P^c(x,v)\\
& P^{c}(x,v) = \frac{1}{Z^{c}} e^{-\frac{1}{2} \beta m v ^2} e^{-\frac{1}{2} \beta k (| x | -x_1)^2}\\
& P^t(x,v) = \frac{1}{Z^c} e^{-\frac{1}{2} \beta m (v -S(x) \dot x_D)^2} e^{-\frac{1}{2} \beta k (|x| -x_D)^2}
\end{align}
with $S(x)$ the sign of $x$.
The PDF $P^\mathrm{sim}(x,v,t)$ is computed from $\num{5e6}$ trajectories, half of them starting from each well, with $X_1=5\sigma_0$ and $\tau=6 f_0^{-1}$. We can then study the relevance of our model by comparing the PDFs. This is done in the movie {\tt PDF\_erasure\_process.mov} (ancillary file), a frame of which is shown as an example in Fig.~\ref{Ansatz} corresponding to $t=4.5f_0^{-1}$. It demonstrates how good is the model to estimate the position and velocity distribution: the oscillations due to the dragging force are replaced by the temperature rise predicted by the compression model. The relaxation in temperature after step 1 predicted by the model also matches the simulation data, except for the transient relaxation oscillations which are not included in the model after the 1st commutation of the system during step 1. The comparison of the PDFs by their ratio in panel (f) also demonstrates that for statistically relevant portion of the phase space, the agreement between the two is better than 20\%. The main deviations occur in the middle of step 1 in the bottom left and top right corner areas. These areas corresponds to trajectories where the system has switched once and presents a mean velocity component from motion of the well it has switched to. The deviation is therefore explained by the fact that the model PDF doesn't includes the mean driving velocity after the first commutation. But because the error made is symmetrical with the initial state, it doesn't impact the computation of average values such as the velocity variance.

\section{\texorpdfstring{$\Teff$}{Effective temperature} approximation} \label{suppmatTeff}

To retrieve the gas analogy,  we apply Eq.~\eqref{eq.W} to the step 1 compression (no deterministic work), and reframe it to identify the volume total derivative: 
\begin{align}
\langle\frac { d\W_1^c}{dt}\rangle&=-  k_B T \frac{\partial \ln \V}{\partial x_1} \dot{x}_1\\
& =-k_BT\frac{d \ln{\V}}{dt}\left(1+\frac{d \ln T}{d \ln (x_1^2/T)}\right)
\end{align}
The second term in the parenthesis on the right hand side can be evaluated from our model once the time evolution of the temperature has been numerically computed. After integration, it represents at most $10\%$ of the final result (upper limit reached for the highest temperature rise in the adiabatic limit). The work required for a fast compression can therefore be approximated by $\langle \W_1^c \rangle \sim k_B \Teff  \ln{2} $, and meet the gas analogy with the effective temperature being worth
\begin{equation}
\Teff=\frac{1}{\ln 2}\int T d\ln{\V}.
\end{equation}
For the erasure cycle in Fig.~\ref{cycle_rapide}, we derive $\Teff=1.35\, T_0$ which gives the compression work with a $6\%$ error. 

\section{Adiabatic limit of the compression work} \label{suppmatadiabiatic}

For large quality factors, heat exchanges with the bath are cancelled:  $d \langle \Q \rangle =0$. Such compressions, called adiabatic compressions (or mean adiabatic~\cite{martinez_adiabatic_2015}), correspond to the highest temperature rise because the kinetic temperature of the system cannot dissipate in the bath. Let us remind that the entropy variation during an adiabatic compression (assumed reversible) vanishes: $ d S = d\langle \Q \rangle/T =0$. Consequently, we have by definition of the entropy: 
\begin{align}
\Delta (k_B \ln{Z^c} + \frac{\langle E^c \rangle}{T})=0\label{DeltaS}
\end{align}
As for $X_1\gg1$ the system starts and ends in the same quadratic potential, the energy in the initial and final states satisfies $\langle E^c_i \rangle /T_0=\langle E^c_f \rangle /T_f=k_B$ (derived from equipartition, or equivalently from $\langle E^c \rangle=\partial \ln{Z^c}/\partial\beta$).

Therefore, only remains in Eq.~\eqref{DeltaS} the variation of the compression partition function written in Eq.~\eqref{Zc}:
\begin{align}
 \Delta (k_B \ln{Z^c} ) = 0 \ \rightarrow \ \Delta(TV)=0
\end{align}
As the volume is divided by two, the temperatures doubles during the adiabatic compression: $T_f=2T_0$. The corresponding work is given by the first law of thermodynamics (with $\langle Q \rangle =0$):
\begin{align}
\langle \W^c \rangle&= \Delta \langle E^c \rangle\\
&=k_BT_f-k_BT_0\\
&=k_BT_0
\end{align}

As a conclusion, an adiabatic compression results in doubling the system temperature and requires on average $k_BT_0$ of work. The above paragraph describes how a quasi-static adiabatic transformation affects the kinetic temperature in response to a volume change. It is worth noticing that Refs.~\onlinecite{martinez_adiabatic_2015, martinez_supplemental_2015} developed the reverse reasoning and prescribe the relevant temperature-volume ratio to be maintained for the realization of adiabatic processes.

\end{document}